\documentclass[nofootinbib,twocolumn,aps,prd,amsmath]{revtex4-2} 
\usepackage{amsmath}
\usepackage{txfonts}
\usepackage{amssymb}
\usepackage{mathrsfs}
\usepackage{graphicx}
\usepackage{epsfig}
\usepackage{dcolumn}
\usepackage{bm}
\usepackage{ulem}
\usepackage{color}

\usepackage{booktabs}    
\usepackage{array}
\usepackage{tabularx}
\usepackage{threeparttable}

\begin{document}

\title{XTE J1814-338 as a strange star admixed with bosonic dark matter}


\author{Shu-Hua Yang$^{1}$}\email{ysh@ccnu.edu.cn}
\author{Fridolin Weber$^2$$^,$$^3$}\email{fweber@ucsd.edu}

\affiliation{$^1$Institute of Astrophysics, Central China Normal
  University, Wuhan 430079, China
\\$^2$Department of Physics, San Diego State
University, San Diego, CA 92182, USA \\$^3$ Department of Physics, University of
 California at San Diego, La Jolla, CA 92093, USA
}


\begin{abstract}
  We show that the compact star XTE J1814–338 can be explained as a
  strange star admixed with self-interacting bosonic dark matter
  (BDM), provided the dark matter fraction exceeds approximately $70\%$. This
  interpretation leads to a robust constraint on the BDM particle
  mass: $m_\chi \lesssim 307(\lambda/\pi)^{1/4}$ MeV ($\lambda$ is the
 dimensionless coupling constant of the BDM). The result is
  independent of formation scenario and microphysical details and is
  falsifiable by future NICER and LIGO/Virgo observations.
\end{abstract}

\maketitle

\noindent {{\it Introduction.}} According to strange quark matter
(SQM) hypothesis \cite{Itoh1970,Bodmer1971,Witten1984,Farhi1984}, SQM
consists of roughly equal numbers of up ($u$), down ($d$), and strange
($s$) quarks, along with a small admixture of electrons, which may be more
stable than ordinary nuclear matter. As a consequence, compact stars
could exist as strange stars (SSs)
\citep{Alcock1986,Madsen1999,Weber2005,Zhang2024} rather than neutron
stars (NSs).Interestingly, it is found recently that SSs could well explain the repeating fast
radio bursts (FRBs) \citep{Geng2021}, and also the anomalous X-ray pulsars (AXPs) and soft gamma-ray repeaters (SGRs) \citep{Qiao2025}.

Compact stars (NSs, SSs) might contain a dark matter (DM) core or a DM halo made of non-annihilating self-interacting DM
\cite{Bramante2024,Grippa2025}.  Thus, it is expected that the
observations of compact stars could help us to unveil the nature of
DM.

Recently, the mass and radius of the Type I X-ray burster and
accretion-powered millisecond pulsar XTE J1814-338 were inferred to be
$M = 1.21 \pm 0.05 M_{\odot}$ and $R = 7.0 \pm 0.4 $ km ($68.3\%$
credibility interval), using the pulse-profile modeling technique
\citep{Kini2024}.  To explain such unusually low mass and radius of
XTE J1814-338, a theoretical model beyond conventional NSs composed of
standard nuclear matter is needed. Several studies have tried to
explain the observations of XTE J1814-338 as hybrid stars
\citep{Laskos-Patkos2025,Zhou2025,Zhang2025}, while others speculate
that XTE J1814-338 might be a DM admixed NS
\citep{Pitz2024,Lopes2025a} or a DM admixed SS
\citep{Yang2025,Lopes2025b}. 

In this Letter, we will explain the observations of XTE J1814-338
assuming it is an SS admixed with bosonic dark matter (BDM)
\citep{Khlopov1985,Colpi1986}.  Although the SSs admixed with BDM have
been studied recently by Liu et al. \citep{Liu2025}, this work will
extend their study to higher values of the fraction of BDM in SSs,
which is necessary for us to explain the mass and radius observations
of XTE J1814-338. Moreover, we will constrain the mass of the BDM
using the observations of XTE J1814-338. This work employs a different
SQM model than Ref.~\citep{Liu2025}.

Compared with neutron or hybrid star models, SSs provide a natural
explanation for compact objects with exceptionally small radii. The
self-bound nature of SQM permits more compact configurations at
relatively low gravitational mass. In this regard, XTE~J1814--338 is
particularly interesting, as its inferred radius of $R = 7.0 \pm 0.4$
km is difficult to reconcile with conventional hadronic equations of
state, but fits well within the predictions of BDM-admixed SSs. This
motivates our investigation within the SS framework, rather than
adopting a nuclear or hybrid star model.

\noindent {\it{Equation of State of SQM and BDM.}} For the
equation of state (EOS) of SQM, we employ the modified MIT bag model
\citep{Farhi1984,Alcock1986,Haensel1986,Weber2005}, where $u$ and $d$
quarks are considered to be massless, while the $s$ quark has a finite
mass ($m_s=93$~MeV \cite{Navas2024}). In this model, first-order
perturbative corrections to the strong interaction coupling constant
$\alpha_{S}$ are included to account for interactions among quarks.
Given the thermodynamic potentials for $u$, $d$, $s$ quarks, and
electrons ($\Omega_{i}$, which can be found in
Refs.~\cite{Alcock1986,Pi2015,Yang2020}), the number density of each species
is
\begin{equation}
n_{i}=-\frac{\partial\Omega_{i}}{\partial\mu_{i}},
\end{equation}
where $\mu_{i}$ ($i=u,d,s,e$) are the chemical potentials. For SQM,
the chemical equilibrium is maintained by the weak-interaction
processes and one has
\begin{equation}
\mu_{d} = \mu_{s} , \quad
\mu_{s} = \mu_{u}+\mu_{e}.
\end{equation}
The charge neutrality equation is given by
\begin{equation}
\frac{2}{3}n_{u}-\frac{1}{3}n_{d}-\frac{1}{3}n_{s}-n_{e}=0.
\end{equation}
The energy density and pressure of SQM are then given by
\begin{eqnarray}
\epsilon_{Q} &=& \sum_{i=u,d,s,e}(\Omega_{i}+\mu_{i}n_{i})+B, \\
p_{Q}        &=& -\sum_{i=u,d,s,e}\Omega_{i}-B,
\end{eqnarray}
respectivley where $B$ is the bag constant.
The EOS of BDM with a repulsive self-interaction is given by
\citep{Colpi1986,Karkevandi2022,Shakeri2024}
\begin{eqnarray}\label{BM1}
p_{D}&=&\frac{m_{\chi}^{4}}{9\lambda}\left(
\sqrt{1+\frac{3\lambda}{m_{\chi}^{4}}\epsilon_{D}}-1\right)^{2}\, ,
\end{eqnarray}
where $\epsilon_{D}$ and $p_{D}$ are the energy density and pressure of BDM, respectively; $m_\chi$ is the BDM particle mass, $\lambda$ is the
dimensionless coupling constant. Derivation of Eq.\ (\ref{BM1}) can be
found in Ref. \citep{Karkevandi2022}. Defining $\epsilon_{0}\equiv
m_{\chi}^{4}/(4\lambda)$, one has \citep{Liu2023,Liu2025}
\begin{eqnarray}\label{BM2}
p_{D}&=&\frac{4\epsilon_{0}}{9}\left(
\sqrt{1+\frac{3}{4\epsilon_{0}}\epsilon_{D}}-1\right)^{2}\, .
\end{eqnarray} 
Apparently, instead of two parameters ($m_\chi$ and $\lambda$), this
EOS only depends on one parameter $\epsilon_{0}$.

For the quartic BDM model used here (the BDM particles are 
described by a complex scalar field with the self-interaction
potential $V(\phi)=\lambda|\phi|^{4}/4$) \citep{Colpi1986,Karkevandi2022}, 
the self-interaction cross section per unit mass is $\sigma/m_\chi =
\lambda^{2}/(16\pi m_\chi^{3})$ in natural units \citep{Liu2023,Liu2025,Eby2016}. For the parameter
range relevant to XTE~J1814--338 ($\lambda=\pi$, $m_\chi \sim
200$--$350$ MeV), this yields $\sigma/m_\chi \sim
10^{-3}\,\mathrm{cm^{2}\,g^{-1}}$, which is orders of magnitude below
astrophysical limits ($\sigma/m_\chi \lesssim
0.1$--$1\ \mathrm{cm^{2}\,g^{-1}}$) \citep{Markevitch2004,Kaplinghat2016,Sagunski2021}.

In this work, we focus on cold catalyzed matter and adopt
zero-temperature EOS for both the SQM and the BDM components. This is
a valid approximation for old, thermally relaxed compact stars such as
XTE J1814–338. For proto-SSs or post-merger remnants,
thermal corrections may become relevant
\cite{Pons1999,Benvenuto1995,Kettner1995,Yasutak2009} and will be
addressed in future work.

\noindent {\it{Results.}} We now derive constraints on the BDM particle mass from the observed mass and
radius of XTE~J1814--338. For the
given EOS of SQM and BDM, we investigate the structure of SSs admixed
with BDM using the two-fluid formalism
\citep{Sandin2009,Li2012,Yang2021,Yang2023,Yang2024,Lu2025,Yang2025}, which means
that we assume the SQM and BDM components interact only through
\begin{figure*}
\includegraphics[width=3.5in]{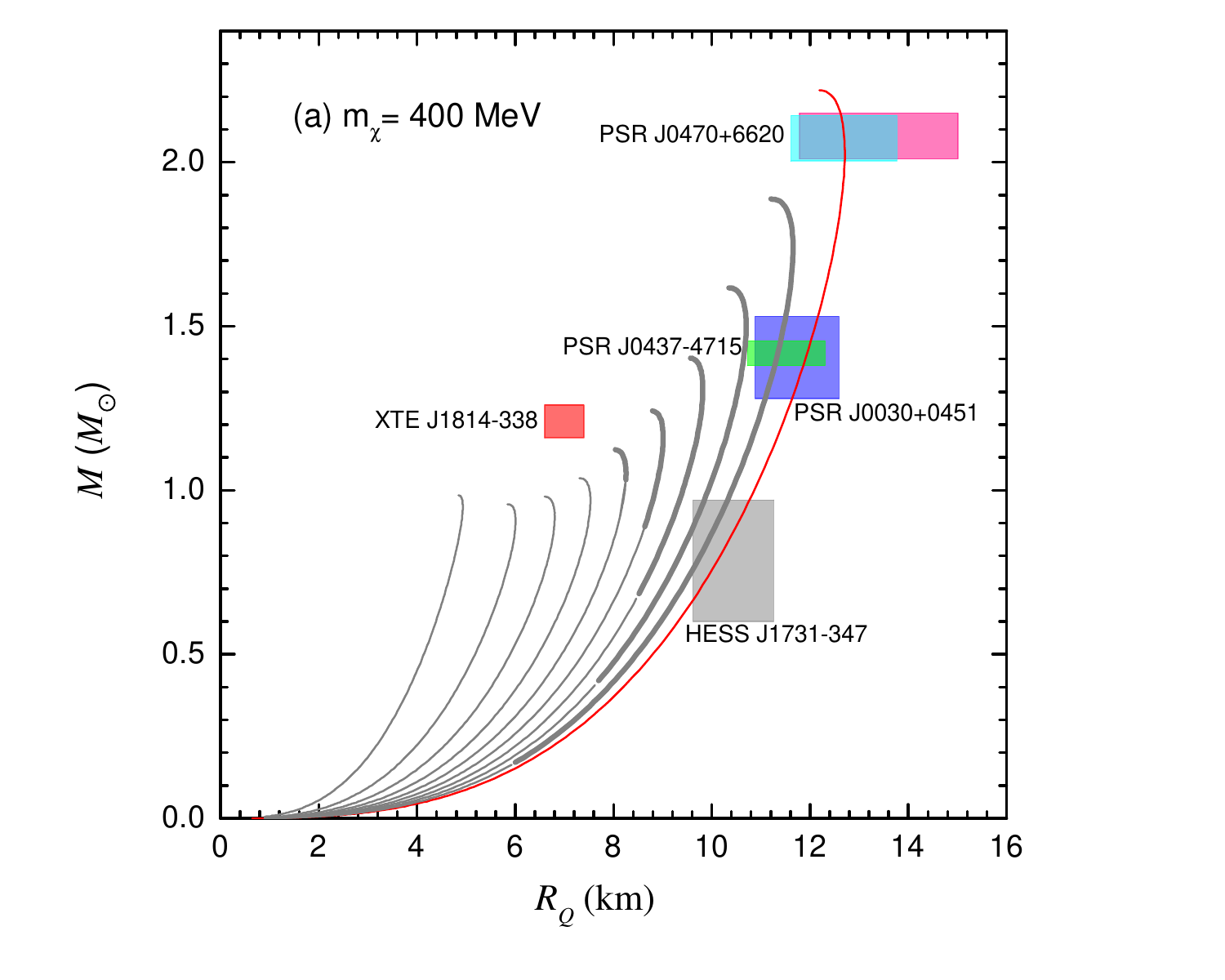}
\includegraphics[width=3.5in]{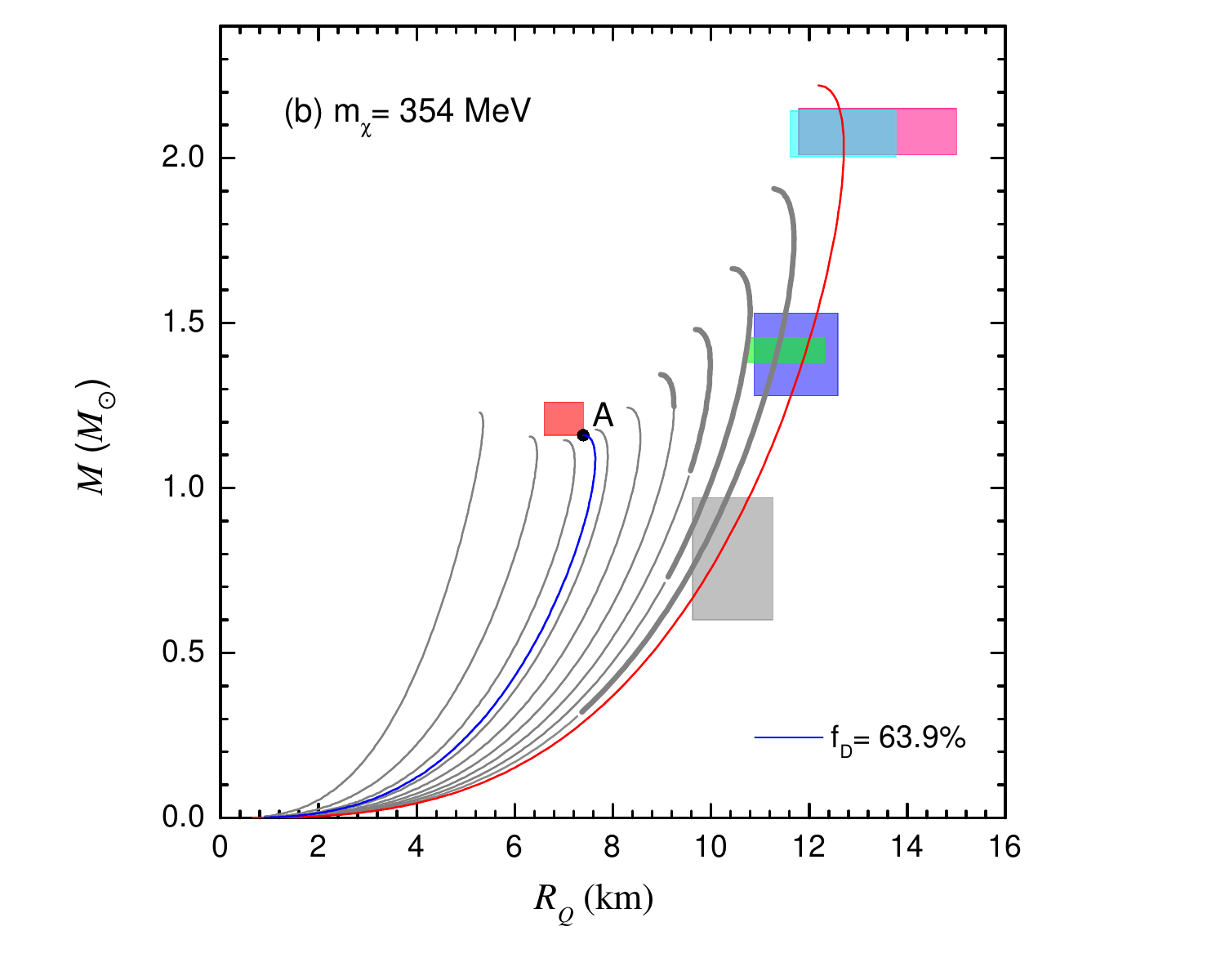}
\includegraphics[width=3.5in]{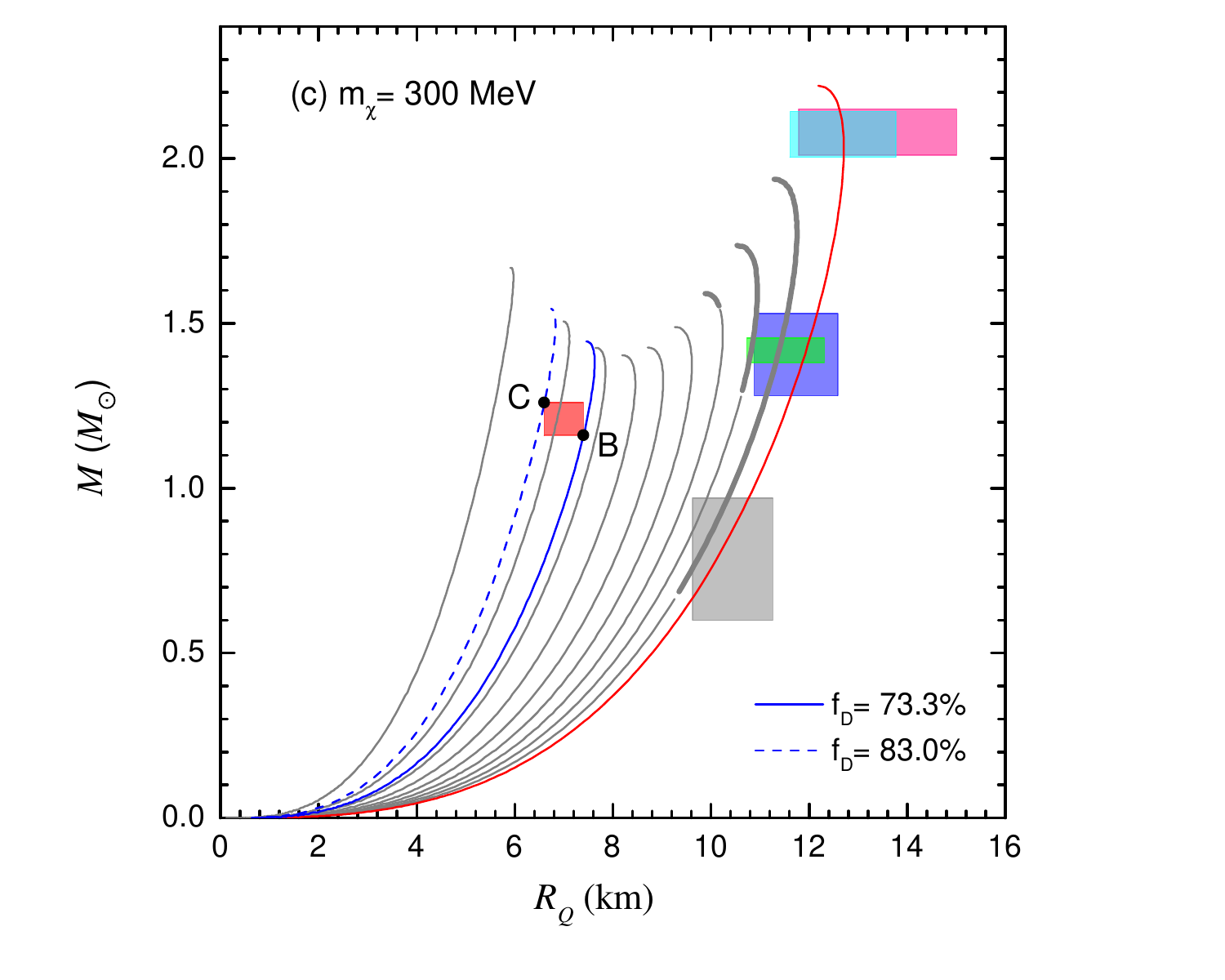}
\includegraphics[width=3.5in]{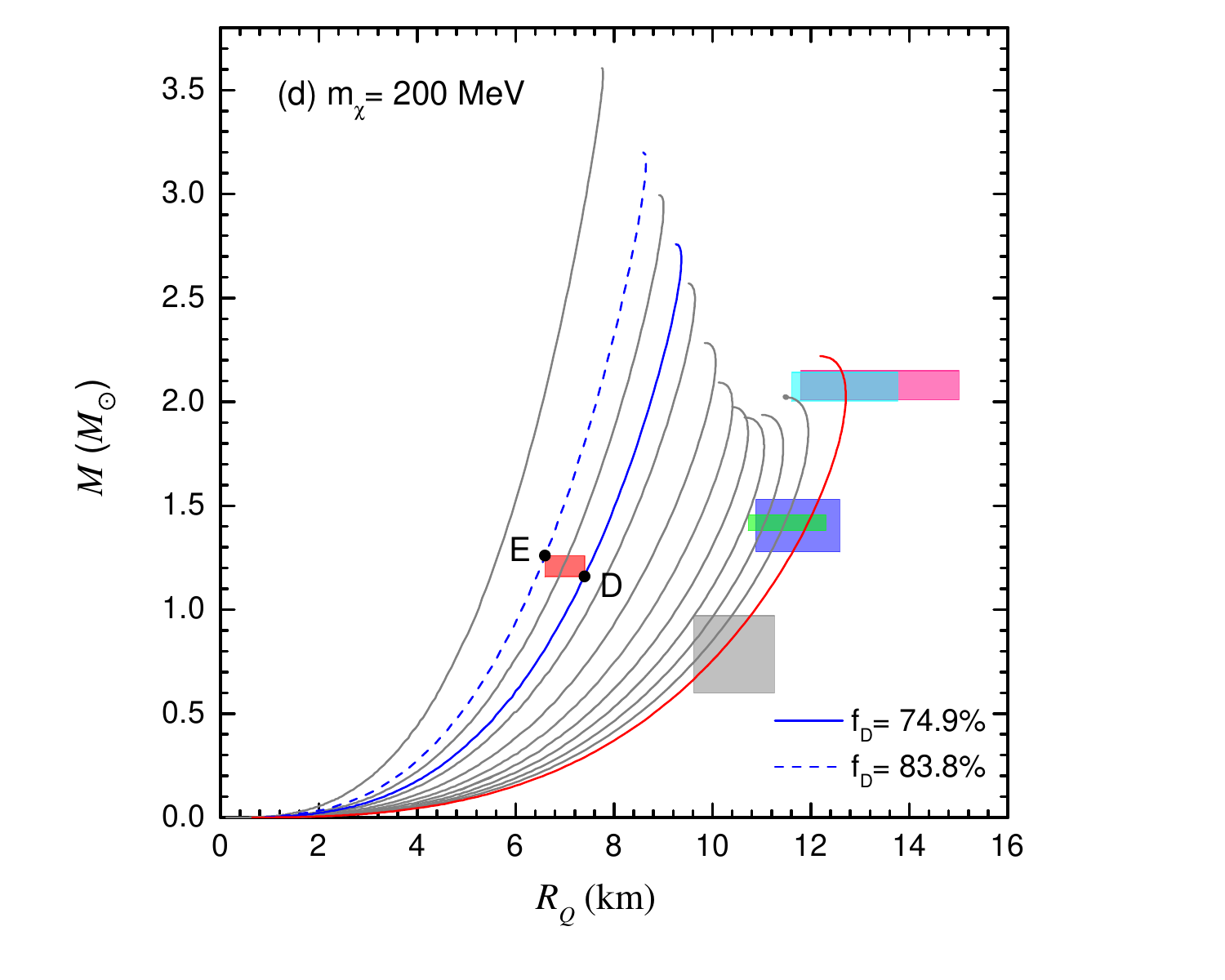}
\caption{ Mass–radius relations for BDM-admixed SSs, with
  $R_Q$ denoting the radius of the SQM component. All models are
  computed using $\alpha_s = 0.6$, $B^{1/4} = 135~\mathrm{MeV}$, and
  $\lambda = \pi$. Panels (a)–(d) correspond to boson masses $m_\chi =
  400$, 354, 300, and 200 MeV, respectively. The red curve in each
  panel shows the pure SS case ($f_D = 0\%$), while gray curves
  represent varying BDM mass fractions from $f_D = 10\%$ to $90\%$ in
  steps of $10\%$. Thick segments correspond to configurations with a DM
  core; thin segments indicate DM halos (a ``DM core'' refers to configurations in which the DM radius does not
exceed the SQM radius, while a ``DM halo'' denotes configurations
with an extended DM distribution surrounding the SQM core). Black dots mark the benchmark
  stellar models listed in Table~\ref{tab:models}. Colored bands show
  observational constraints from: XTE~J1814--338
  (red)~\cite{Kini2024}, PSR~J0740+6620
  (pink/cyan)~\cite{Cromartie2020,Fonseca2021,Dittmann2024,Salmi2024},
  PSR~J0030+0451 (blue)~\cite{Vinciguerra2024}, PSR~J0437--4715
  (green)~\cite{Choudhury2024}, and HESS~J1731--347
  (gray)~\cite{Doroshenko2022}.  }
\label{rmDASS}
\end{figure*}
gravity, with no direct interactions between them. In our calculation,
the mass fraction of BDM is defined as $f_{D} \equiv M_{D}/M$, where
$M$ is the total mass of the star and $M_{D}$ is the mass of the BDM
component.

As shown in Fig.\ \ref{rmDASS}(a), the mass and radius of XTE J1814-338 cannot be
satisfied when $m_{\chi}=400$ MeV. However, the curves move upward as
the value of $m_{\chi}$ becomes smaller, and the observations of XTE
J1814-338 can be marginally satisfied for the case of $m_{\chi}=354$
MeV, supposing the BDM mass fraction of the star is $f_{D}= 63.9\%$,
see Fig.\ \ref{rmDASS}(b). If $m_{\chi}< 354$ MeV, the observations of
XTE J1814-338 could be well satisfied for certain ranges of the value
of $f_{D}$. For example, for the case of $m_{\chi}= 300$ MeV, the
BDM-admixed SSs with $f_{D}= 73.3\%-83.0\%$ can explain the
\begin{table*}
\caption{Parameters of five BDM-admixed SS models that
  satisfy the mass and radius constraints of XTE~J1814--338. $R_D$ is
  the total radius of the star (i.e., the outer radius of the BDM halo), $M(R_Q)$ is the enclosed gravitational mass at the
  surface of the SQM core, and $M_c$ is the BDM mass outside
  $R_Q$.}
\label{tab:models}
\centering
\begin{ruledtabular}
\begin{tabular}{lccccccccc}
Model & $m_\chi$ (MeV) & $f_D$ (\%) & $M$ ($M_\odot$) & $R_Q$ (km) & $M_D$ ($M_\odot$) & $R_D$ (km) & $M(R_Q)$ ($M_\odot$) & $M_c$ ($M_\odot$) & $M_c/M_D$ \\
\hline
A & 354 & 63.9 & 1.16 & 7.40 & 0.741 & 9.19 & 1.096 & 0.064 & 0.086 \\
B & 300 & 73.3 & 1.16 & 7.40 & 0.850 & 18.03 & 0.566 & 0.594 & 0.699 \\
C & 300 & 83.0 & 1.26 & 6.60 & 1.046 & 19.81 & 0.403 & 0.857 & 0.820 \\
D & 200 & 74.9 & 1.16 & 7.40 & 0.869 & 48.18 & 0.320 & 0.840 & 0.967 \\
E & 200 & 83.8 & 1.26 & 6.60 & 1.056 & 52.81 & 0.222 & 1.038 & 0.983 \\
\end{tabular}
\end{ruledtabular}
\end{table*}
observations of XTE J1814-338 (Fig.\ \ref{rmDASS}(c)). For another
example, for the case of $m_{\chi}=200$ MeV, the BDM-admixed SSs with
$f_{D}= 74.9\%-83.8\%$ can explain the observations of XTE J1814-338
(Fig.\ \ref{rmDASS}(d)). Note that to fulfill the observations of XTE
J1814-338, the BDM mass fraction should be as high as $f_D \sim 70\%$.
In contrast, low values of $f_D$ (less than $\sim 30\%$, even pure SS
with $f_D=0$) could satisfy all the other observations shown in
Fig.\ \ref{rmDASS}.  This distinguishes XTE J1814–338 as a uniquely
high-$f_D$ object among observed compact stars, whereas others are
consistent with low-$f_D$ or pure SS configurations.

The parameters of the five stellar models marked by black dots in
panels (b), (c), and (d) are shown in Table \ref{tab:models}. We find that
stellar model "A" has a compact halo, and the other four stellar
models ("B", "C", "D", and "E") have an intermediate halo
\cite{Shawqi2024}, where a ``DM halo'' denotes configurations
with an extended DM distribution surrounding the SQM core. Different from Ref. \cite{Shawqi2024} with a
small DM fraction ($f_{D}=5\%$), $f_{D}$ is large for all the
stellar models in Table \ref{tab:models}. As a result, one cannot use the
value of $M_c/M_D$ ($M_c$ is the BDM mass located outside
  $R_Q$, where $R_Q$ denotes the radius of the SQM component, i.e. the visible surface) alone to distinguish the type of halos (compact,
intermediate, or diffuse).  In Ref. \cite{Shawqi2024}, the authors
demonstrated that for the scenario of the diffuse halos, the mass
measurement of XTE J1814-338 (which is determined through X-ray
pulse-profile modeling \citep{Kini2024}) corresponds to $M(R_Q)$ (the
enclosed gravitational mass at $R_Q$), while for the scenarios of the
compact halos and the intermediate halos, the mass measurement of XTE
J1814-338 will be very close to the total mass of the star
($M$). Therefore, it is more reasonable to choose $M$ as the
longitudinal axis in Fig.\ \ref{rmDASS} rather than $M(R_Q)$. As for
the radius, since photons are emitted from the surface of the SQM, the
visible radius inferred from X-ray pulse-profile modeling corresponds
to $R_Q$, the radius of the baryonic component. We therefore treat
$R_Q$ as the observable radius throughout this work, in contrast to
the total gravitational radius $R_D$ of the full DM halo.

Our work leads to the conclusion that for $\alpha_{S}=0.6$,
$B^{1/4}=135$ MeV, and $\lambda=\pi$, the observations of XTE
J1814-338 constrain the boson mass of the BDM to $m_{\chi}\leq 354$
MeV. Since $\epsilon_{0}= m_{\chi}^{4}/(4\lambda$) is the only
parameter which the EOS of BDM depends on, the constraint to
$m_{\chi}$ turns out to be $m_{\chi}\leq 354 (\lambda /\pi)^{1/4}$
MeV. This constraint could be reframed in terms of the compactness
$\mathcal{C}_{\rm obs} = M/R_Q$ inferred from
observation. Specifically, XTE~J1814--338 exhibits a compactness of
$\mathcal{C}_{\rm obs} \approx 1.21\,M_\odot / 7.0\,\text{km} \approx
0.173\,M_\odot/\text{km}$. We analyze the models in Table~\ref{tab:models} and
interpolate their corresponding compactness values as a function of
the BDM EOS parameter $\epsilon_0$. By extrapolating this relation,
one finds that only models with $\epsilon_0 \lesssim 7.1 \times
10^8\,\text{MeV}^4$ can satisfy the compactness constraint.
This yields a general upper bound,
\begin{equation}
m_\chi \lesssim 307 \left( \frac{\lambda}{\pi} \right)^{1/4}~\mathrm{MeV} .
\end{equation}
While this constraint is derived from XTE~J1814–338, it is phrased in
terms of compactness and applies to any compact object with similar
mass–radius ratios, regardless of composition or formation
history. Although the bound still depends on the existence of an
extremely compact star, it does not rely on detailed assumptions about
internal structure or DM distribution. Rather, it reflects a
minimal requirement that any BDM+SS configuration must satisfy to
reproduce such compactness.

We verified that these BDM-admixed SS configurations lie on the dynamically
stable branch of the two-fluid sequence, well below the turning point at which
$dM/d\rho_c$ changes sign.\footnote{In multifluid stars, the onset of instability is associated with the turning point of the
gravitational mass along the equilibrium sequence (e.g., Ref. \cite{Prix2004}). For the one-parameter
sequences shown in Fig.~1, this turning point occurs at the maximum mass, so all models lying
on the $dM/d\rho_c>0$ side of this maximum belong to the stable branch.}

\noindent {\it{Outlook.}}  To place our
results in the broader landscape of DM admixed compact star
models, we compare the viability of SSs mixed with BDM, as presented
in this work, against several existing frameworks, including NSs and
SSs with similar dark sector
interactions. Table~\ref{tab:comparison} summarizes the extent
to which each model can explain the mass and radius constraints of
well-observed compact objects, such as XTE~J1814--338, PSR~J0740+6620,
PSR~J0030+0451, PSR J0437-4715, and HESS~J1731--347.

We note that interpretations across models depend sensitively on the
assumed BDM fraction $f_D$, the boson mass $m_\chi$, and which
observational constraints are prioritized. For example, Karkevandi et
al.~\cite{Karkevandi2022} require that both the 2\,$M_\odot$ mass
constraint and the GW170817 tidal deformability bound $\Lambda_{1.4}
\lesssim 580$ \cite{Abbott2018} be satisfied with the same value of
$f_D$, leading them to reject models with $f_D \gtrsim 5\%$. However,
\begin{table*}
\caption{Comparison of compact star observations with different
  BDM-admixed models. “Viable” means the model satisfies mass and
  radius constraints under the authors' assumptions. For Karkevandi et
  al.~\cite{Karkevandi2022}, this requires $f_D \lesssim 5\%$ due to
  tidal deformability constraints. For Liu et al.~\cite{Liu2023},
  high-$f_D$ NSs are disfavored for $m_\chi \gtrsim 200$
  MeV.}
\label{tab:comparison}
\centering
\begin{ruledtabular}
\begin{tabular}{lcccc}
Object & Karkevandi et al.~\cite{Karkevandi2022} & Liu et al.~\cite{Liu2023} & Liu et al.~\cite{Liu2025} & This work \\
       & (NS+BDM) & (NS+BDM) & (SS+BDM) & (SS+BDM) \\
\hline
XTE J1814--338       & Not viable ($f_D \lesssim 5\%$)       & Viable ($f_D \gtrsim 5\%$ plausible) & --                            & Viable ($m_\chi \leq 354$ MeV, $f_D \gtrsim 64\%$) \\
PSR J0740+6620       & Viable                                & Viable                              & Viable                       & Viable \\
PSR J0030+0451       & Viable                                & Viable                              & Viable                       & Viable \\
PSR J0437--4715      & Viable                                & Viable                              & Viable                       & Viable \\
HESS J1731--347      & --                                     & --                                   & Viable (DM-core SS)         & Viable \\
PSR J5014--4002E     & --                                     & --                                   & Viable (DM-halo SS)         & Viable (pure SS, $M_{\max} = 2.22\,M_\odot$) \\
\end{tabular}
\end{ruledtabular}
\end{table*}
as we argue, different compact objects may naturally have different
values of $f_D$, depending on their formation history. Similarly, Liu
et al.~\cite{Liu2023} find that high-$f_D$ NS configurations are
disfavored for $ m_\chi \gtrsim 200 \,\mathrm{MeV}$, but this does not
rule out high-$f_D$ SSs. In our work, the high mass of PSR
J5014--4002E  \citep{Barr2024} is naturally accommodated by a pure SS model with $M_{\rm
  max} = 2.22\,M_\odot$.
Here, we want to mention that Liu et al. \cite{Liu2023} have studied
BDM-admixed NSs using the same EOS of BDM as that in this work. From
Fig. 7 in Ref.~\cite{Liu2023}, one finds that the mass and radius
observations of XTE J1814-338 could also be explained by BDM-admixed
NSs, and the parameter of the BDM model could also be constrained.

Note that although the mass and radius measurements of XTE J1814--338 could also be consistent with BDM-admixed NSs (BDM--NS), BDM-admixed SS interpretation offers a more robust explanation
for the extremely small radius inferred for XTE~J1814--338. SS matter is self-bound and naturally supports radii in the $6$--$8$ km range even at low masses,
whereas obtaining such radii within BDM--NS models typically requires fine-tuning
of both the hadronic EOS and the DM fraction. In our framework, the SS
interpretation reproduces the observed compactness of XTE~J1814--338 with fewer
modeling assumptions.

Achieving DM mass fractions as high as $f_D \sim 70\%$ cannot
be explained by standard accretion from the Galactic halo, which
contributes at most $\sim 10^{-5}\,M_\odot$ over a stellar
lifetime~\cite{Sandin2009,Bramante2024}. More plausible scenarios
involve the early formation of compact, self-gravitating DM
cores—such as from primordial overdensities or DM-rich collapse in
regions of enhanced DM density (e.g., subhalo remnants or DM
spikes near black holes). These configurations could later accrete
baryons and evolve into SSs with high $f_D$. Similar ideas
have been explored in the context of dark
stars~\cite{Spolyar2008,Ilie2023} and two-fluid DM–admixed
NSs~\cite{Kumar2024}. Self-interactions, as modeled via the coupling
$\lambda$, may further promote core formation by suppressing
free-streaming and enhancing clumping.  Although a full cosmological
model is beyond the scope of this work, high-$f_D$ configurations
remain plausible and warrant further exploration.

We stress that such high values of $f_D$ only appear under special circumstances mentioned above and they are \emph{not}
expected for ordinary stars since standard halo accretion contributes at most $\sim 10^{-5}\,M_\odot$ over
a stellar lifetime. Therefore, our model does \emph{not} affect current understanding of stellar structure and evolution, nor does it conflict with current cosmological models. Also our model does \emph{not} imply that most pulsars contain substantial DM fractions. As shown in Fig.~1, only the remarkably small radius of XTE~J1814--338 appears to be exceptional, all other well-measured compact stars
(PSR~J0740+6620, PSR~J0030+0451, PSR~J0437--4715, HESS~J1731--347) may be explained with
$f_D \lesssim 30\%$, and in many cases $f_D = 0$.

Beyond mass–radius fits and tidal deformability bounds, future
observations could distinguish BDM-admixed SSs through spectral and
dynamical signatures. NICER and ATHENA are sensitive to redshift,
surface gravity, and light bending—all of which are enhanced by the
steep potentials of high-$f_D$ stars. In the gravitational wave
sector, post-merger remnants of NS–SS or SS–SS mergers may exhibit
delayed collapse, altered f-mode spectra, or broader emission due to
the extended DM halo. These features fall within the detection
thresholds of LIGO A+, Einstein Telescope, and Cosmic Explorer. Thus,
BDM-admixed compact stars offer not only a compelling explanation for
compact objects like XTE~J1814–338, but also falsifiable predictions
for multimessenger astronomy.

Finally, we note that our two-fluid formalism assumes only
gravitational coupling between the SQM and BDM sectors. Future work
should investigate the effects of weak or self-annihilating
interactions, which could impact stability, cooling, and structure.

In summary, XTE J1814–338 provides a rare observational window into
the structure of DM–admixed compact stars. We demonstrate
that a high-$f_D$ BDM-admixed SS model not only matches the measured
mass and radius, but also constrains the BDM particle
mass to $m_\chi \lesssim 307(\lambda/\pi)^{1/4}~\mathrm{MeV}$. This
offers a testable prediction for future high-precision radius
measurements and gravitational wave observations.

\noindent{\it{Acknowledgments.}} The authors thank professor Renxin Xu
for discussions on the mass measurement of XTE J1814-338. This work is
supported by the National Key R\&D Program of China (Grant
No.\ 2021YFA0718504).

\bibliographystyle{apsrev4-1} 
\bibliography{references}

\end{document}